\documentclass[reprint, aps, prl, floatfix,superscriptaddress]{revtex4-2}
\usepackage{graphicx}    
\usepackage{dcolumn}    
\usepackage{bm}             
\usepackage{amssymb}   
\usepackage{amsmath}    
\usepackage{amsthm}
\usepackage{mathrsfs}    
\usepackage{commath}   
\usepackage{subfigure}    
\usepackage{braket}
\usepackage{color}
\usepackage{siunitx}
\usepackage[colorlinks=true, allcolors=blue]{hyperref}
\usepackage{hyphenat}
\usepackage{footnote}
\usepackage{notes2bib}
\usepackage{xcolor}
\newcommand{\beq}{\begin{equation}}
\newcommand{\eeq}{\end{equation}}

\begin{document}

\title{Hardware-native quantum phase estimation with circuit QED}

\author{Changchun Zhong}
\email{zhong.changchun@xjtu.edu.cn}

\affiliation{Department of Physics, Xi'an Jiaotong University, Xi'an, Shaanxi 710049, China}

\date{\today}

\begin{abstract}
Quantum phase estimation is a cornerstone algorithm for determining eigenvalues of unitary operators with Heisenberg-limited precision. Conventional implementations rely on digital controlled-unitary operations together with phase-extraction circuits, which generally results in substantial circuit depth and hardware overhead. Here, we propose a hardware-native alternative that replaces digital controlled-unitary operations by analog bosonic interactions, naturally available in circuit quantum electrodynamics. The protocol extracts the phase through a sequence of binary threshold tests. A bosonic mode serves as an efficient quantum memory where the binary digits of the phase are encoded into the direction of phase-space rotations. These digits are then read out sequentially via high-fidelity homodyne measurements. We show that the protocol preserves Heisenberg scaling in estimation precision while simultaneously providing exponentially suppressed failure probability. By exploiting bosonic degrees of freedom and analog dispersive interactions, the scheme provides a hardware-efficient realization of quantum phase estimation and establishes a natural route toward implementing high-precision phase estimation on circuit quantum electrodynamics platforms.
\end{abstract}

\maketitle

\textit{Introduction}---The prospect of solving certain computational problems more efficiently than any known classical algorithm is a primary driving force behind the development of quantum computers. A prominent example is Shor's factoring algorithm \cite{shor1994}, which can factor large integers using resources that scale polynomially with the problem size. By contrast, the best known classical factoring algorithms exhibit sub-exponential complexity. This exponential separation highlights the transformative potential of quantum computation and has stimulated extensive efforts toward realizing scalable quantum hardware \cite{eisert2025,monroe2013,bluvstein2026}.


As a fundamental subroutine underlying many quantum algorithms, quantum phase estimation (QPE) plays a central role in a wide range of quantum applications, including quantum chemistry, materials science, quantum simulation, and quantum machine learning \cite{aspuru2005,babbush2018,harrow2009,roushan2017,biamonte2017}. Consider a system Hamiltonian $\hat{H}_s$ satisfying
$
\hat{H}_s\ket{\psi_k}=\lambda_k\ket{\psi_k},
$
where, without loss of generality, the spectrum is assumed to lie within the interval $\mathrm{sp}(\hat{H}_s)\in[0,1)$ after an appropriate rescaling and energy shift. Standard QPE (s-QPE) estimates the eigenvalues $\lambda_k$ by applying controlled powers of the unitary evolution
$
U=e^{-i\hat{H}_s t/\hbar},
$
followed by an inverse quantum Fourier transform (QFT). Since the eigenenergy is encoded as the phase accumulated under the unitary evolution, the procedure is often referred to as quantum phase estimation \cite{watrous2018}.

Operationally, QPE implements the transformation
\begin{equation}
\sum_k c_k\ket{\psi_k}\otimes\ket{\mathrm{anc}}
\xrightarrow{\mathrm{QPE}}
\sum_k c_k\ket{\psi_k}\otimes\ket{\lambda_k'},
\end{equation}
where $\ket{\mathrm{anc}}$ denotes an $n$-qubit ancilla register and $\ket{\lambda_k'}$ encodes an $n$-bit approximation of the eigenvalue $\lambda_k$. The estimation error decreases exponentially with the register size, yielding a precision
$
\abs{\lambda_k^\prime-\lambda_k}\le\epsilon\sim \mathcal{O}(2^{-n}).
$
Consequently, s-QPE requires only $\mathcal{O}(n)$ ancilla qubits and $\mathcal{O}(n^2)$ gates for the inverse QFT, while a constant number of repetitions suffices to achieve a fixed confidence level. Furthermore, s-QPE attains the Heisenberg limit in the sense that the total coherent evolution time scales as
$T\sim\mathcal{O}(1/\epsilon)$, which is asymptotically optimal for phase estimation \cite{hayashi2014}.

\begin{figure*}[t]
\centering
\includegraphics[width=\textwidth]{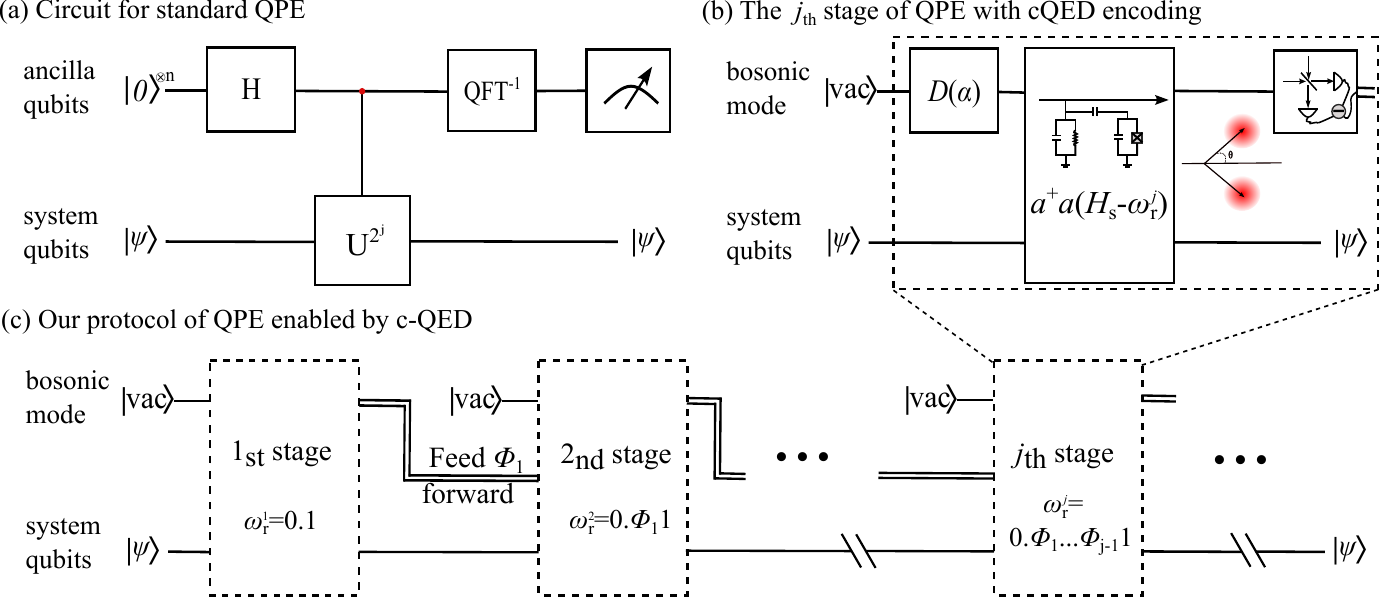}
\caption{(a) Quantum circuit for standard QPE. (b) The $j_\text{th}$ stage of our QPE protocol with c-QED encoding and homodyne measurement. A bosonic mode is used to replace the ancilla qubits in standard QPE or Kitaev's iterative QPE. The protocol is assisted by well-developed toolbox from c-QED. (c) Schematics for our QPE protocol enabled by binary threshold test and adaptive frequency tuning. The test result is sequentially feed forward to extract the phase information bit by bit, which is encoded in the mode frequency as $\omega_r^j=0.\phi_1\phi_2...\phi_j...$ (in binary fraction). \label{fig1}}
\end{figure*}

\begin{table*}[t]
\caption{Comparison of the QPE protocols } 
\label{tab1}
\begin{center}
\begin{tabular}{c|c|c|c|c|c|c}
\hline
\hline
Protocol & Ancilla & Inverse QFT   & Controlled U$^{2^j}$ gates  & Feedback  & Hardware  & Heisenberg scaling\\
\hline
Standard QPE   & $n$ qubits &  Yes   & Yes  &   No & Digital    & Yes \\
Kitaev's i-QPE   & $1$ qubit &  No   & Yes  &   Qubit rotation & Digital    & Yes\\
Our protocol   & $1$ bosonic mode &  No   & No  &   Frequency tuning & Analog  & Yes  \\
\hline
\hline
\end{tabular}
\end{center}
\end{table*}

In practice, however, s-QPE relies on a deep controlled implementation of the unitary evolution together with an inverse QFT. Realizing these operations through Trotterization, block encoding, or qubitization \cite{berry2015,low2019} generally incurs significant circuit-depth and hardware overhead \cite{wiebe2016,russo2021,gu2023}. As a result, the resource requirements of practical implementations may substantially exceed the ideal Heisenberg-limited scaling. This challenge has motivated the development of alternative phase-estimation protocols. A prominent example is Kitaev's iterative QPE (i-QPE) \cite{kitaev1995}, which replaces the inverse QFT and large ancilla register with a single ancilla qubit, extracting the phase bits sequentially through adaptive qubit rotations and classical feedback.


In this work, we propose a conceptually simple and experimentally practical protocol for QPE. The central idea is to replace digital controlled-unitary operations and adaptive qubit manipulations by analog dispersive evolution of a bosonic mode, thereby providing a hardware-native realization of i-QPE, as illustrated in Fig.~\ref{fig1}. The bosonic mode is dispersively coupled to the target system through the widely used cQED Hamiltonian
$
\hat{H}=\chi \hat{a}^{\dagger}\hat{a}\left(\hat{H}_s-\omega_r\right),
$
where $\hat{a}$ is the annihilation operator of the bosonic mode, $\chi$ denotes the dispersive coupling strength, and $\omega_r$ is the tunable mode frequency \cite{blais2021}. Under this interaction, the eigenvalue information of $\hat{H}_s$ is continuously mapped onto the phase-space evolution of the bosonic mode. To attain the Heisenberg-limited precision, the phase is extracted iteratively through a sequence of binary threshold tests. Each test determines a single binary digit of the eigenvalue by appropriately tuning the mode frequency $\omega_r$, a capability that is readily available in state-of-the-art cQED architectures. The binary outcomes are obtained via homodyne measurements of the bosonic mode, closely resembling the high-fidelity dispersive readout techniques routinely employed for superconducting qubits \cite{blais2021}. We show that the measurement overhead contributes only logarithmically to the overall resource cost. Consequently, the protocol preserves the optimal scaling
$T=\mathcal{O}\left(\frac{1}{\epsilon}\log\frac{1}{\delta}\right)$ in the precision $\epsilon$ while simultaneously providing an exponentially suppressed failure probability $\delta$. 

In superconducting platforms, high-Q cavities naturally provide long-lived bosonic memories with high-fidelity dispersive readout and frequency tunability. The proposed protocol exploits these native capabilities directly, avoiding the synthesis of controlled powers of the target unitary through digital gate sequences. To highlight the differences, a comparison among different QPE is given in Tab.~\ref{tab1}.

\textit{Protocol description}---Our protocol consists of $n$ iterative stages, as shown in Fig.~\ref{fig1}(c). Each stage performs a binary decision that progressively narrows the interval containing the unknown eigenvalue. The decision threshold is determined by the frequency of the bosonic mode. For simplicity, consider an initial product state
$
\ket{\alpha}\otimes\ket{\psi},
$
where $\ket{\alpha}$ is a coherent state of the bosonic mode and $\ket{\psi}$ is an eigenstate of the system Hamiltonian $\hat{H}_s$ with eigenvalue $\lambda$ (note extending to superposition of multiple eigenstates $\ket{\psi}=\sum_k c_k\ket{\psi_k}$ is straightforward and we use single eigenstate to convey the main idea). During the $j_\text{th}$ stage, the system evolves under the dispersive interaction $\hat{H}_j=\chi\hat{a}^{\dagger}\hat{a}\left(\hat{H}_s-\omega_r^{j}\right)$,
where the tunable mode frequency $\omega_r^{j}$ serves as the threshold value for the binary decision (its explicit choice will be specified below). Since $\ket{\psi}$ is an eigenstate of $\hat{H}_s$, the system Hamiltonian may be replaced by its eigenvalue $\lambda$, yielding the effective Hamiltonian
\begin{equation}\label{eqhj}
\hat{H}_j=\chi\hat{a}^{\dagger}\hat{a}\Delta_j
\end{equation}
with $\Delta_j=\lambda-\omega_r^{j}$. The resulting dynamics correspond to a phase-space rotation of the coherent state. The direction of rotation is determined solely by the sign of $\Delta_j$: the state rotates clockwise for $\Delta_j>0$ and counterclockwise for $\Delta_j<0$. Therefore, measuring the rotation direction reveals whether the eigenvalue $\lambda$ lies above or below the threshold $\omega_r^{j}$. By adaptively updating the threshold at each stage, the protocol successively determines the binary digits of $\lambda$, thereby producing an $n$-bit estimate of the eigenvalue.



Denoting the eigenvalue in binary form as $\lambda=0.\phi_1\phi_2\cdots=\sum_{l=1}^{\infty}\phi_l2^{-l}$
with $\phi_l\in{0,1}$, the threshold frequency for the $j$th stage is chosen as
\begin{equation}\label{eqmd}
\omega_r^j=
\sum_{l=1}^{j-1}\phi_l2^{-l}
+2^{-j},
\qquad j=1,\ldots,n.
\end{equation}
Namely, $\omega_r^j$ corresponds to the midpoint of the interval determined by the previously extracted bits. Since the digits $\phi_l$ are initially unknown, they are inferred sequentially from earlier measurements and subsequently used to determine the threshold frequency for the next stage, as illustrated in Fig.~\ref{fig1}(c). 

The algorithm can be summarized as follows:
\begin{enumerate}
\item \textit{Initialization}.---Prepare the bosonic mode in a coherent state $\ket{\alpha}$ with real positive amplitude $\alpha$, and prepare the system in an eigenstate $\ket{\psi}$ of the Hamiltonian $\hat{H}_s$ whose eigenvalue $\lambda$ is to be estimated.

\item \textit{Evolution}.---For the $j_\text{th}$ stage, set the mode frequency to $\omega_r^j$ according to Eq.~(\ref{eqmd}) and evolve the system under the interaction Hamiltonian $\hat{H}_j$. The evolution time is chosen as $t_j\sim\mathcal{O}(2^j)$.

\item \textit{Measurement}.---Perform a homodyne measurement of the bosonic mode to determine the sign of
$
\Delta_j=\lambda-\omega_r^j.
$
The measurement outcome reveals whether $\lambda$ lies above or below the threshold and therefore determines the binary digit $\phi_j$.

\item \textit{Iteration}.---Using the newly acquired digit $\phi_j$, update the threshold frequency $\omega_r^{j+1}$ and repeat the previous steps. After $n$ stages, the protocol outputs the estimate
$
\lambda' = 0.\phi_1\phi_2\cdots\phi_n,
$
with precision $\epsilon=\mathcal{O}(2^{-n})$.
\end{enumerate}


\begin{figure*}[t]
\centering
\includegraphics[width=\textwidth]{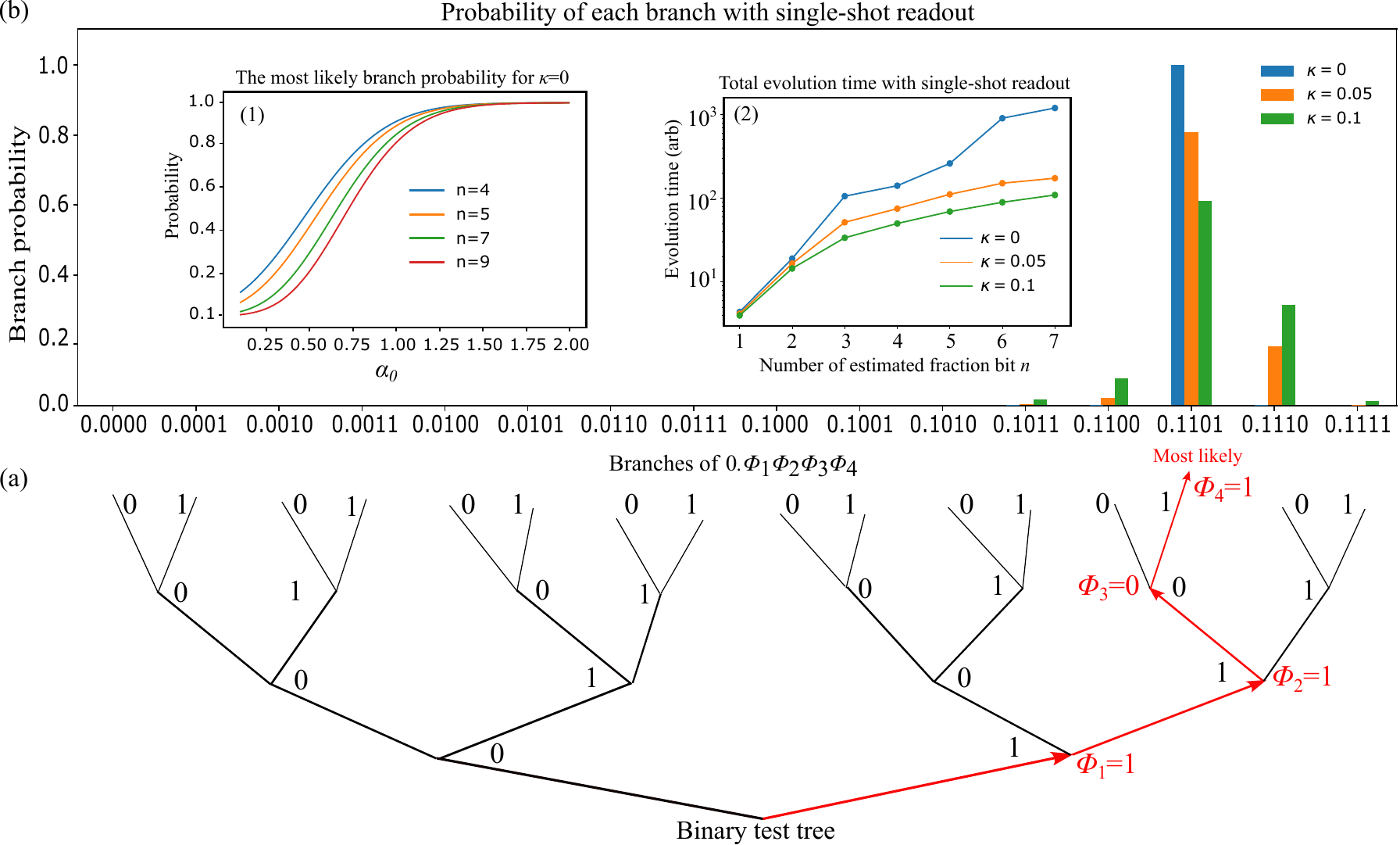}
\caption{The numerical example using our QPE protocol to estimate an eigenvalue. The true eigenvalue is chosen as $\lambda=0.11011011101$. We take arbitrary units and set the coupling $\chi=1$. All branches are realized by single-shot readout at each stage. (a) The binary test tree, with the most probable path depicted in red. The path traces the best estimate of the eigenvalue. (b) The probability for all possible branches, up to the fourth fraction digits. The histogram shows that the branch $0.\phi_1\phi_2\phi_3\phi_4=0.1101$, matching the true eigenvalue, is the most probable. In this example, we choose $\ket{\alpha_0=2.0}$ and the mode loss $\kappa=\{0.0,0.05,0.1\}$. The legend plot (1) inside shows the probability for the most probable branches as a function of the initial $\alpha_0$ with $\kappa=0$, where the probability approaches one when the initial mode occupation number gets larger for varied precision. The legend plot (2) shows the total evolution time of the most likely branch with respect to the precision. The case with $\kappa=0$ directly shows the agreement with the Heisenberg scaling $T\sim\mathcal{O}(1/\epsilon)$, as shown by the blue line. The orange and green lines with $\kappa>0$ bend down since the bosonic mode evolves to the maximal separation with less time at each stage, and it corresponds to a reduced success branch probability which can be increased by repeating the process \label{fig2}}
\end{figure*}


\textit{Performance analysis}---After $n$ stages, the protocol determines the first $n$ binary digits of the eigenvalue,
$
\lambda' = 0.\phi_1\phi_2\cdots\phi_n,
$
yielding an estimation error
$
\abs{\lambda^\prime-\lambda} = \mathcal{O}(2^{-n})$. To analyze the required evolution time, consider the $j_\text{th}$ stage. Under the effective Hamiltonian (\ref{eqhj}), the coherent state undergoes a phase-space rotation by an angle
$
\theta_j=\chi\Delta_j t_j,
$
where $\Delta_j=\lambda-\omega_r^j$. Since the threshold frequency $\omega_r^j$ is chosen to match the first $(j-1)$ binary digits of $\lambda$, the residual detuning is determined by the unresolved lower-order bits and therefore scales as
$\Delta_j\sim\mathcal{O}(2^{-j})$. The sign of $\Delta_j$ determines the direction of rotation, while its magnitude determines the angular separation between the two candidate phase-space distributions. To maintain a constant discrimination fidelity at every stage, the rotation angle $\theta_j$ must remain of order unity. Consequently, the interrogation time scale as $t_j\sim\mathcal{O}(2^j)$. Thus, the total evolution time $T\sim\sum_{j=1}^{n}2^j=2(2^n-1)$. Using $\epsilon=\mathcal{O}(2^{-n})$, we obtain $T\sim\mathcal{O}\left(\frac{1}{\epsilon}\right)$, which corresponds to Heisenberg-limited scaling and is asymptotically optimal for phase estimation.


The binary threshold test at each stage is closely related to the dispersive readout of a superconducting qubit in cQED \cite{blais2021,swiadek2024,hatridge2013}. For each measurement round, the homodyne detector is configured to measure the momentum quadrature of the bosonic mode. The measurement outcome follows a Gaussian distribution whose mean depends on the sign of $\Delta_j$. Ideally, a negative outcome indicates $\Delta_j>0$ (equivalently, $\phi_j=1$), whereas a positive outcome indicates $\Delta_j<0$ (equivalently, $\phi_j=0$). In practice, however, the finite width of the Gaussian distribution gives rise to a nonzero probability that the measurement outcome crosses the decision threshold, resulting in an incorrect bit assignment. To further suppress the overall failure probability, the homodyne measurement may be repeated multiple times.

In order to analyze the effect with repeated measurements, we consider the following model where we denote the single shot success probability as $p$, and define the random variable 
\begin{equation}
q=
    \begin{cases}
        +1, \text{probability } p, \\
        -1, \text{probability } 1-p.
    \end{cases}
\end{equation}
The sampling average $\hat{q}=\frac{1}{m}\sum_k^m q_k$ and the true mean is given as $\mu=2p-1$. According to the Hoeffding inequality, $\hat{q}$ gives a good estimation of the mean with the failure probability bound \cite{rohatgi2015}
\begin{equation}
    \text{Pr}(|\hat{q}-\mu|\ge\xi)\le 2e^{-2m\xi^2}.
\end{equation}
If the objective were to estimate the mean $\mu$ with accuracy $\xi$, then achieving a failure probability at most $\delta$ would require the repetition
$
m\sim\mathcal{O}\left(\frac{1}{\xi^2}\log\frac{1}{\delta}\right),
$
which corresponds to the familiar standard-quantum-limit.
 
However, the goal is not to estimate $\hat{q}$ accurately. Instead, we only need to distinguish between two hypotheses $\Delta_j>0$ and $\Delta_j<0$, depending solely on the sign of the sample average $\hat{q}$. Thus the threshold is $\hat{q}_\text{treshold}=0$, and the gap is $\abs{\hat{q}_\text{threshold}-\mu}=2p-1$, which is a constant. The error probability becomes 
\begin{equation}
    \text{Pr}_\text{err}(|\hat{q}-\mu|\ge 2p-1)\le 2e^{-2m(2p-1)^2},
\end{equation}
which gives the scaling $m\sim\mathcal{O}(\frac{1}{(2p-1)^2}\log \frac{1}{\delta})$. We see the small quantity $\xi$ is replaced by a constant. It can be understood as: for the threshold test to fail, one would need the sample average $\hat{q}$ to be more than $2p-1$ distance away from the true mean $\mu$, which is very improbable. Since this overhead is on top of the time scaling, the overall complexity is thus derived as
\begin{equation}
    T\sim\mathcal{O}(\frac{1}{\epsilon}\log\frac{1}{\delta}),
\end{equation}
which is asymptotically optimal in both the estimation precision and the failure probability.

\textit{Examples with loss}---Consider a system prepared in a single eigenstate of $\hat{H}_s$ with the corresponding true eigenvalue $\lambda=0.11011011101$ and the task is to estimate this eigenvalue up to the $n_\text{th}$ binary fraction digit. We initialize the bosonic mode in a coherent state $\ket{\alpha_0\in\mathbb{R}^+}$. At the $j_\text{th}$ stage, the bosonic mode evolves under the Hamiltonian Eq.~(\ref{eqhj}) with a fixed mode loss $\kappa$, result in $\ket{\alpha(t_j)}=\ket{\alpha_0e^{-(i\chi\Delta_j+\frac{\kappa}{2})t_j}}$. A homodyne measurement is then performed to get the momentum quadrature $\mathcal{P}_j=(a-a^\dagger)/i\sqrt{2}$. The result is positive with the probability
\begin{equation}\label{succpro}
    \text{Pr}(\mathcal{P}_j>0)=\frac{1}{2}(1+\text{Erf}(\Bar{\mathcal{P}}_j)),
\end{equation}
where $\Bar{\mathcal{P}}_j=\sqrt{2}\text{Im}(\alpha(t_j))$. Obviously, the discrimination error is minimized when the coherent state undergoes a phase-space rotation, which maximizes the separation of the two hypotheses along the measured quadrature. If $\kappa=0$, the separation is maximized when the rotation angle reaches $\pi/2$ and the interrogation time is $t_j\sim\pi/2\chi\Delta_j$, while for finite loss the rotation angle would be smaller and can be achieved with less time (discuss later). Depending on the sign of the measurement outcome, the threshold frequency is updated according to Eq.~(\ref{eqmd}) for the following stage. 

Eventually, a binary tree grows and the most probable branch gives the estimate of the eigenvalue, as shown in Fig.~\ref{fig2}(a). For demonstration, we present the numerical results based on \textit{single-shot} readout for each stage. The probability of each branch can factorize according to the sequential structure of the protocol, e.g., Pr$(\phi_1\phi_2\phi_3\phi_4)=\text{Pr}(\phi_1)\text{Pr}(\phi_2|\phi_1)\text{Pr}(\phi_3|\phi_1\phi_2)\text{Pr}(\phi_4|\phi_1\phi_2\phi_3)$. The histogram in Fig.~\ref{fig2}(b) shows the probabilities of all branches up to the first four binary digits of the eigenvalue, with the dominant path $\phi_1=1\rightarrow\phi_2=1\rightarrow\phi_3=0\rightarrow\phi_4=1$, in agreement with the true eigenvalue. The probability depends sensitively both on the mode loss and the initial coherent state amplitude, as shown by the varied parameters in the histogram and the legend plot (1). The reason is similar to that in dispersive qubit readout---reduced photon number increases single-shot error rates, thereby suppressing the weight of the correct branch. The degradation can be mitigated either by increasing the initial coherent-state amplitude throughout the protocol or by repeating the homodyne measurement at each stage, both of which introduce an overhead that is captured by the failure probability parameter $\delta$.

The influence of mode loss is more pronounced when estimating higher-order binary digits because the evolution time grows with the stage index. The legend plot (2) shows the total evolution time associated with the most probable branch as a function of the precision ($n$-bits). For $\kappa=0$, the blue curve exhibits the expected exponential growth, consistent with the Heisenberg-limited scaling $T\sim\mathcal{O}(1/\epsilon)$. In contrast, the orange and green curves for $\kappa>0$ bend downward, reflecting a shorter optimal interrogation time. At first sight, this reduction appears to surpass the Heisenberg limit. In reality, however, it originates from cavity loss: if the loss $\kappa=0$, the optimal interrogation time is $t_j=\pi/(2\chi\Delta_j)$; for non-zero loss, however, coherent-state damping shifts the optimal operating point to an earlier time 
\begin{equation}
t_j=\arccos({\kappa}/{\sqrt{\kappa^2+4\chi^2\Delta^2_j}})/(\chi\Delta_j),
\end{equation}
which maximizes the phase-space separation between the two hypotheses. This shorter evolution time is accompanied by a reduced single-shot success probability. Recovering the original confidence therefore requires a constant number of repeated threshold measurements, which increases the total evolution time while preserving the asymptotic Heisenberg scaling. Finite mode loss therefore does not fundamentally limit the protocol. Instead, it modifies the optimal interrogation time without changing the overall computational complexity.

\textit{Discussion}---State-of-the-art superconducting cavities routinely achieve quality factors exceeding $10^{10}$ \cite{rosenblum2018,mirrahimi2016}, corresponding to photon lifetimes approaching one second at microwave frequencies, while dispersive couplings of $\chi/2\pi\sim 1-5$ MHz have been demonstrated in cQED \cite{wallraff2004,vlastakis2013,blais2021,hazra2025,touzard2019,hatridge2013}. For a representative example with the stage $j=15$, $\chi/2\pi=2$ MHz and $\kappa=10^3$ Hz, the optimal interrogation time is reduced from $4.1$ ms in the lossless limit to $1.7$ ms. Under these parameters, Eq.~(\ref{succpro}) predicts a single-shot success probability of approximately $80\%$, compared with $99\%$ for $\kappa=0$. By repeating the threshold test only a constant number of times with a simple majority vote, the original confidence level can be substantially while preserving the asymptotic complexity of the protocol.  

It is worth noting that finite homodyne efficiency and amplifier noise likewise reduce the single-shot discrimination fidelity. Nevertheless, existing cQED platforms routinely achieve single-shot readout fidelity exceeding $99.9\%$ \cite{blais2021,walter2017,hazra2025,touzard2019,hatridge2013}. Together with the robustness of the binary-threshold protocol and the favorable asymptotic scaling established above, these experimental capabilities indicate that high-precision quantum phase estimation is compatible with current superconducting architectures. More broadly, our work demonstrates that bosonic degrees of freedom can provide a hardware-native platform for implementing quantum algorithms. By replacing digital controlled-unitary operations with analog dispersive interactions while preserving the optimal asymptotic scaling, the proposed protocol establishes a practical route toward hardware-efficient quantum phase estimation and related quantum-information processing tasks.

\begin{acknowledgments}
We acknowledge the funding support from Xi'an Jiaotong University (Grant No. 11301224010717), Shaanxi Fundamental Science Research Project for Mathematics and Physics (Grant No. 25JSY006), and the Youth Scientist funding support from Shaanxi Province (Grant No. 2024SYJ21).
\end{acknowledgments}

\bibliography{all}

\onecolumngrid



\end{document}